\documentclass[12pt]{amsart}
\usepackage{amssymb}
\usepackage[russian]{babel}
\usepackage{graphicx}
\newcommand{\bea}{\begin{eqnarray}}
\newcommand{\eea}{\end{eqnarray}}
\newcommand{\ba}{\begin{array}}
\newcommand{\ea}{\end{array}}
\newcommand{\bc}{\begin{center}}
\newcommand{\ec}{\end{center}}

\newcommand{\be}{\begin{equation}}
\newcommand{\ee}{\end{equation}}

\begin{document}
\sloppy
%\Large
%\Large
╙─╩ 517.98

\textbf{╨.╠.╒ръшьют} (╚эёЄшЄєЄ ьрЄхьрЄшъш, ╥р°ъхэЄ,
╙чсхъшёЄрэ.) \\

\textbf{╤╦└┴╬ ╧┼╨╚╬─╚╫┼╤╩╚┼ ╠┼╨█ ├╚┴┴╤└ ─╦▀ ═╤-╠╬─┼╦╚ ─╦▀
═╬╨╠└╦▄═╬├╬ ─┼╦╚╥┼╦▀ ╚═─┼╩╤└ ╫┼╥█╨┼}\\

In this paper is studied HC-model on a Cayley tree and under some
conditions on parameters of the HC-model we prove existence of the
weakly periodic (non periodic) Gibbs measures for normal divisor
of the index four.\\

┬ фрээющ ёЄрЄ№х шчєўрхЄё  ═╤-ьюфхы№ эр фхЁхтх ╩¤ыш. ╧Ёш эхъюЄюЁ√ї
єёыютш ї эр ярЁрьхЄЁ√ яюърчрэю, ўЄю ёє∙хёЄтє■Є ёырсю яхЁшюфшўхёъшх (эх яхЁшюфшўхёъшх) ьхЁ√ ├шссёр фы  эюЁьры№эюую фхышЄхы  шэфхъёр ўхЄ√Ёх.\\

\textbf{╩ы■ўхт√х ёыютр}: фхЁхтю ╩¤ыш, ъюэЇшуєЁрЎш , ═╤-ьюфхы№,
ьхЁр ├шссёр, яхЁшюфшўхёъшх ьхЁ√, ёырсю яхЁшюфшўхёъшх ьхЁ√.\\

\textbf{1. ┬тхфхэшх.} ╧юэ Єшх ьхЁ√ ├шссёр шуЁрхЄ трцэє■ Ёюы№ т
ёЄрЄшёЄшўхёъющ ьхїрэшъх. ╬ёэютэющ яЁюсыхьющ фрээюую урьшы№Єюэшрэр
 ты хЄё  юяшёрэшх тёхї юЄтхўр■∙шї хьє ьхЁ ├шссёр. ╬яЁхфхыхэшх ьхЁ√
├шссёр ш фЁєушї яюэ Єшщ, ёт чрээ√ї ё ЄхюЁшхщ ьхЁ ├шссёр, ьюцэю
эрщЄш, эряЁшьхЁ, т ЁрсюЄрї \cite{6}-\cite{Si}. ─ы  ьюфхыш ╚чшэур
эр фхЁхтх ╩¤ыш ¤Єр чрфрўр шчєўхэр фюёЄрЄюўэю яюыэюёЄ№■. ╥ръ,
эряЁшьхЁ, т ЁрсюЄх \cite{Bl} яюёЄЁюхэю эхёўхЄэюх ьэюцхёЄтю ъЁрщэшї
ушссёютёъшї ьхЁ, р т ЁрсюЄх \cite{Bl1} эрщфхэю эхюсїюфшьюх ш
фюёЄрЄюўэюх єёыютшх ъЁрщэюёЄш эхєяюЁ фюўхээющ Їрч√ ьюфхыш ╚чшэур
эр фхЁхтх ╩¤ыш.

┬ ьэюуюўшёыхээ√ї ЁрсюЄрї (ёь. эряЁшьхЁ \cite{1}-\cite{7}) эр
фхЁхтх ╩¤ыш шчєўхэ√ яхЁшюфшўхёъшх ьхЁ√ ├шссёр фы  Ёрчышўэ√ї
ьюфхыхщ ёЄрЄшёЄшўхёъющ ьхїрэшъш. ▌Єш ьхЁ√, т юёэютэюь, с√ыш
ЄЁрэёы Ўшюээю-шэтрЁшрэЄэ√ьш, ышсю яхЁшюфшўхёъшьш ё яхЁшюфюь фтр.
┴юыхх Єюую, фы  ьэюушї ьюфхыхщ эр фхЁхтх ╩¤ыш фюърчрэю, ўЄю
ьэюцхёЄтю яхЁшюфшўхёъшї ьхЁ ├шссёр юўхэ№ схфэю, Є.х., ёє∙хёЄтє■Є
{\it Єюы№ъю} яхЁшюфшўхёъшх ушссёютёъшх ьхЁ√ ё яхЁшюфюь фтр (ёь.
эряЁшьхЁ \cite{3}-\cite{Ro}).

╫Єюс√ яюыєўшЄ№ сюыхх °шЁюъюх ьэюцхёЄтю ушссёютёъшї ьхЁ, т ЁрсюЄрї \cite{8}, \cite{9} ттхфхэ√ сюыхх
юс∙шх яюэ Єш  яхЁшюфшўхёъющ ьхЁ√ ├шссёр, Є.х. ёырсю яхЁшюфшўхёъшх ушссёютёъшх ьхЁ√ ш фюърчрэю
ёє∙хёЄтютрэшх Єръшї ьхЁ фы  ьюфхыш ╚чшэур эр фхЁхтх ╩¤ыш.

┬ ЁрсюЄх \cite{7} шчєўхэр HC (Hard Core)-ьюфхы№ эр фхЁхтх ╩¤ыш ш
фюърчрэр, ўЄю ЄЁрэЎы Ўшюээю-шэтрЁшрэЄэр  ьхЁр ├шссёр фы  ¤Єющ
ьюфхыш хфшэёЄтхээр. ╩Ёюьх Єюую, яЁш эхъюЄюЁ√ї єёыютш ї эр
ярЁрьхЄЁ√ ═╤-ьюфхыш фюърчрэр эххфшэёЄтхээюёЄ№ яхЁшюфшўхёъшї ьхЁ
├шссёр ё яхЁшюфюь фтр.

┬ ЁрсюЄх \cite{RKh} шчєўхэ√ ёырсю яхЁшюфшўхёъшх ьхЁ√ ├шссёр фы 
HC-ьюфхыш фы  эюЁьры№эюую фхышЄхы  шэфхъёр фтр ш яЁш эхъюЄюЁ√ї
єёыютш ї эр ярЁрьхЄЁ√ яюърчрэр хфшэёЄтхээюёЄ№
(ЄЁрэёы Ўшюээю-шэтрЁшрэЄэюёЄ№) ёырсю яхЁшюфшўхёъющ ьхЁ√ ├шссёр, р
т ЁрсюЄх \cite{XR} яюърчрэр хфшэёЄтхээюёЄ№
(ЄЁрэёы Ўшюээю-шэтрЁшрэЄэюёЄ№) ёырсю яхЁшюфшўхёъющ ьхЁ√ ├шссёр фы 
HC-ьюфхыш яЁш ы■с√ї чэрўхэш ї ярЁрьхЄЁют.

▌Єр ЁрсюЄр яюёт ∙хэр шчєўхэш■ ёырсю яхЁшюфшўхёъшї ьхЁ ├шссёр
═╤-ьюфхыш фы  эюЁьры№эюую фхышЄхы  шэфхъёр ўхЄ√Ёх эр фхЁхтх ╩¤ыш ш
фюърч√трхЄё  ёє∙хёЄтютрэшх ёырсю яхЁшюфшўхёъшї (эх яхЁшюфшўхёъшї)
ьхЁ ├шссёр эр эхъюЄюЁ√ї шэтрЁшрэЄрї яЁш эхъюЄюЁ√ї єёыютш ї
эр ярЁрьхЄЁ√.\\

\textbf{2. ╧ЁхфтрЁшЄхы№э√х ётхфхэш .} ─хЁхтю ╩¤ыш $\tau^k$
яюЁ фър $ k\geq 1 $ - схёъюэхўэюх фхЁхтю, Є.х. уЁрЇ схч Ўшъыют, шч
ърцфющ тхЁ°шэ√ ъюЄюЁюую т√їюфшЄ Ёютэю $k+1$ ЁхсхЁ. ╧єёЄ№
$\tau^k=(V,L,i)$, уфх $V$ --- хёЄ№ ьэюцхёЄтю тхЁ°шэ $\tau^k$, $L$
--- хую ьэюцхёЄтю ЁхсхЁ ш $i$ --- ЇєэъЎш  шэЎшфхэЄэюёЄш,
ёюяюёЄрты ■∙р  ърцфюьє ЁхсЁє $l\in L$ хую ъюэЎхт√х Єюўъш $x, y \in
V$. ┼ёыш $i (l) = \{ x, y \} $, Єю $x$ ш $y$ эрч√тр■Єё   {\it
сышцрщ°шьш ёюёхф ьш тхЁ°шэ√} ш юсючэрўрхЄё  $l = \langle
x,y\rangle $. ╨рёёЄю эшх $d(x,y), x, y \in V$ эр фхЁхтх ╩¤ыш
юяЁхфхы хЄё  ЇюЁьєыющ
$$
d (x, y) = \min \{d | \exists x=x_0,x_1, \dots, x_{d-1},
x_d=y\in V \ \ \mbox {Єръшх, ўЄю} \ \ \langle x_0,x_1\rangle,\dots, \langle x_
{d-1}, x_d\rangle\}.$$

─ы  ЇшъёшЁютрээюую $x^0\in V$ юсючэрўшь $$ W_n = \ \{x\in V\ \ | \
\ d (x, x^0) =n \}, \  V_n = \ \{x\in V\ \ | \ \ d (x, x^0) \leq n
\}$$

─ы  $x\in W_{n}$ юсючэрўшь
$$ S(x)=\{y\in{W_{n+1}}:d(x,y)=1\}.$$

 ╧єёЄ№ $\Phi=\{0,1\}$ ш
$\sigma\in\Phi^V$-ъюэЇшуєЁрЎш , Єю хёЄ№ $\sigma=\{\sigma(x)\in
\Phi: x\in V\}$, уфх $\sigma (x)=1$ ючэрўрхЄ, ўЄю тхЁ°шэр $x$ эр
фхЁхтх ╩¤ыш чрэ Єр , р $\sigma (x)=0$ ючэрўрхЄ, ўЄю юэр ётюсюфэр .

╩юэЇшуєЁрЎш  $\sigma$ эрч√трхЄё  фюяєёЄшьющ, хёыш $\sigma
(x)\sigma (y)=0$ фы  ы■с√ї ёюёхфэшї $\langle x,y \rangle $ шч $V
 (V_n $ шыш $W_n$, ёююЄтхЄёЄтхээю ) ш юсючэрўшь ьэюцхёЄтю Єръшї
ъюэЇшуєЁрЎшщ ўхЁхч $\Omega$ ($\Omega_{V_n}$ ш $\Omega_{W_n}).$
▀ёэю, ўЄю $\Omega\subset\Phi^V.$

├рьшы№Єюэшрэ HC-ьюфхыш юяЁхфхы хЄё  яю ЇюЁьєых
  $$H(\sigma)=\left\{%
\begin{array}{ll}
    J \sum\limits_{x\in{V}}{\sigma(x),} \ \ \ $ хёыш $ \sigma \in\Omega $,$ \\
   +\infty ,\ \ \ \ \ \ \ \ \ \ $  \ хёыш $ \sigma \ \notin \Omega $,$ \\
\end{array}%
\right. $$ уфх $J\in R$.

╧єёЄ№ $\mathbf{B}-$  $\sigma$- рыухсЁр, яюЁюцфхээр 
ЎшышэфЁшўхёъшьш яюфьэюцхёЄтрьш $\Omega.$ ─ы  ы■сюую $n$ юсючэрўшь
ўхЁхч $\mathbf{B}_{V_n}=\{\sigma\in\Omega:
\sigma|_{V_n}=\sigma_n\}$ яюфрыухсЁє $\mathbf{B},$ уфх
$\sigma|_{V_n}-$ ёєцхэшх $\sigma$ эр $V_n,$ $\sigma_n: x\in V_n
\mapsto \sigma_n(x)-$ фюяєёЄшьр  ъюэЇшуєЁрЎш  т $V_n.$

\textbf{╬яЁхфхыхэшх 1}. ─ы  $\lambda >0$ ═╤-ьхЁр ├шссёр хёЄ№
тхЁю ЄэюёЄэр  ьхЁр $\mu$ эр $(\Omega , \textbf{B})$ Єрър , ўЄю фы 
ы■сюую $n$ ш $\sigma_n\in \Omega_{V_n}$
$$
\mu \{\sigma \in \Omega:\sigma|_{V_n}=\sigma_n\}=
\int_{\Omega}\mu(d\omega)P_n(\sigma_n|\omega_{W_{n+1}}),
$$
уфх
$$
P_n(\sigma_n|\omega_{W_{n+1}})=\frac{e^{-H(\sigma_n)}}{Z_{n}
(\lambda ; \omega |_{W_{n+1}})}\textbf{1}(\sigma_n \vee \omega
|_{W_{n+1}}\in\Omega_{V_{n+1}}).
$$

╟фхё№ ёшьтюы $\vee$ ючэрўрхЄ юс·хфшэхэшх ъюэЇшуєЁрЎшщ ш $Z_n
(\lambda ; \omega|_{W_{n+1}})$-- эюЁьшЁютюўэ√щ ьэюцшЄхы№ ё
уЁрэшўэ√ь єёыютшхь $\omega|_{W_n}$:
$$
Z_n (\lambda ; \omega|_{W_{n+1}})=\sum_{\widetilde{\sigma}_n \in
\Omega_{V_n}}
e^{-H(\widetilde{\sigma}_n)}\textbf{1}(\widetilde{\sigma}\vee
\omega|_{W_{n+1}}\in \Omega_{V_{n+1}}).
$$

╚чтхёЄэю \cite{1}, \cite{2}, ўЄю $\tau^k$ ьюцэю яЁхфёЄртшЄ№ ъръ
$G_k$ - ётюсюфэюх яЁюшчтхфхэшх  $k+1$ Ўшъышўхёъшї уЁєяя тЄюЁюую
яюЁ фър ё юсЁрчє■∙шьш $a_1,...,a_{k+1}$, ёююЄтхЄёЄтхээю.

╧єёЄ№ $\widehat{G}_k-$ яюфуЁєяяр уЁєяя√ $G_k$.

┼ёыш ушссёютёър  ьхЁр шэтрЁшрэЄэр юЄэюёшЄхы№эю эхъюЄюЁющ яюфуЁєяя√
 ъюэхўэюую шэфхъёр $ \widehat{G}_k\subset {G_k}$, Єю юэр эрч√трхЄё 
 $ \widehat{G}_k$ - яхЁшюфшўхёъющ.

╚чтхёЄэю \cite{7}, ўЄю ърцфющ ьхЁх ├шссёр фы  HC-ьюфхыш эр фхЁхтх
╩¤ыш ьюцэю ёюяюёЄрты Є№ ёютюъєяэюёЄ№ тхышўшэ $z=\{z_x, x\in G_k
\},$ єфютыхЄтюЁ ■∙шї
$$z_x=\prod_{y \in S(x)}(1+\lambda z_y)^{-1}, \eqno (1)$$ уфх $\lambda=e^J>0$-
ярЁрьхЄЁ.\

\textbf{╬яЁхфхыхэшх 2}. ╤ютюъєяэюёЄ№ тхышўшэ $z=\{z_x,x\in G_k\}$
эрч√трхЄё  $ \widehat{G}_k$-яхЁшюфшўхёъющ, хёыш  $z_{yx}=z_x$ фы 
$\forall x\in G_k, y\in\widehat{G}_k.$\

$G_k-$ яхЁшюфшўхёъшх ёютюъєяэюёЄш эрч√тр■Єё  ЄЁрэёы Ўшюээю-шэтрЁшрэЄэ√ьш.

─ы  ы■сюую $x\in G_k $ ьэюцхёЄтю $\{y\in G_k: \langle
x,y\rangle\}\setminus S(x)$ шьххЄ хфшэёЄтхээ√щ ¤ыхьхэЄ, ъюЄюЁюую
юсючэрўшь ўхЁхч $x_{\downarrow}$ (ёь.\cite{5},\cite{8}).

╧єёЄ№ $G_k/\widehat{G}_k=\{H_1,...,H_r\}$ ЇръЄюЁ уЁєяяр, уфх
$\widehat{G}_k$ эюЁьры№э√щ фхышЄхы№ шэфхъёр $r\geq 1.$\

\textbf{╬яЁхфхыхэшх 3}. ╤ютюъєяэюёЄ№ тхышўшэ $z=\{z_x,x\in G_k\}$
эрч√трхЄё  $\widehat{G}_k$ - ёырсю яхЁшюфшўхёъющ, хёыш
$z_x=z_{ij}$ яЁш $x\in H_i, x_{\downarrow}\in H_j$ фы  $\forall
x\in G_k$.\

╟рьхЄшь, ўЄю ёырсю яхЁшюфшўхёър  ёютюъєяэюёЄ№ $z$ ёютярфрхЄ ё
юс√ўэющ яхЁшюфшўхёъющ (ёь. юяЁхфхыхэшх 2), хёыш чэрўхэшх $z_x$ эх
чртшёшЄ юЄ $x_{\downarrow}$.\

\textbf{╬яЁхфхыхэшх 4}. ╠хЁр $\mu$ эрч√трхЄё 
$\widehat{G}_k$-(ёырсю) яхЁшюфшўхёъющ, хёыш юэр ёююЄтхЄёЄтєхЄ
$\widehat{G}_k$-(ёырсю) яхЁшюфшўхёъющ ёютюъєяэюёЄш тхышўшэ $z$.\\

\textbf{3. ╤ырсю яхЁшюфшўхёъшх ьхЁ√ ├шссёр.} ╧єёЄ№
$H_{\{a_{1}\}}=\{x\in G_k: w_x(a_1)-\mbox{ўхЄэюх ўшёыю}\},$ уфх
$w_x(a_1)-$ ўшёыю сєът√ $a_1$ т ёыютх $x\in G_k$,
${G_k}^{(2)}=\{x\in G_k: \mid x\mid-\mbox{ўхЄэюх ўшёыю}\},$ уфх
$\mid x\mid-$ фышэр ёыютр $x\in G_k$ ш
${G_k}^{(4)}=H_{\{a_{1}\}}\cap{G_k}^{(2)}-$ ёююЄтхЄёЄтє■∙шщ хьє
эюЁьры№э√щ фхышЄхы№ шэфхъёр 4.\

╨рёёьюЄЁшь ЇръЄюЁ-уЁєяяє ${G_k}^{(4)}=\{H_0, H_1, H_2, H_3\},$ уфх
$$H_0=\{x\in G_k: w_x(a_1)-\mbox{ўхЄэю}, |x|-\mbox{ўхЄэю}\}$$
$$H_1=\{x\in G_k: w_x(a_1)-\mbox{эхўхЄэю}, |x|-\mbox{ўхЄэю}\}$$
$$H_2=\{x\in G_k: w_x(a_1)-\mbox{ўхЄэю}, |x|-\mbox{эхўхЄэю}\}$$
$$H_3=\{x\in G_k: w_x(a_1)-\mbox{эхўхЄэю}, |x|-\mbox{эхўхЄэю}\}$$
╥юуфр т ёшыє (1) $G_k-$ ёырсю яхЁшюфшўхёър  ёютюъєяэюёЄ№ $z_x$
шьххЄ тшф
$$
z_x=\left\{%
\begin{array}{ll}
    z_1, & {x \in H_3, \ x_{\downarrow} \in H_1} \\
    z_2, & {x \in H_1, \ x_{\downarrow} \in H_3} \\
    z_3, & {x \in H_3, \ x_{\downarrow} \in H_0} \\
    z_4, & {x \in H_0, \ x_{\downarrow} \in H_3} \\
    z_5, & {x \in H_1, \ x_{\downarrow} \in H_2} \\
    z_6, & {x \in H_2, \ x_{\downarrow} \in H_1} \\
    z_7, & {x \in H_2, \ x_{\downarrow} \in H_0} \\
    z_8, & {x \in H_0, \ x_{\downarrow} \in H_2}, \\
    \end{array}%
\right.$$ уфх $z_x$ єфютыхЄтюЁ хЄ ёшёЄхьх єЁртэхэшщ
$$
\left\{%
\begin{array}{ll}
    z_{1}=\frac{1}{(1+\lambda z_4)^i}\cdot\frac{1}{(1+\lambda z_2)^{k-i}} \\
    z_{2}=\frac{1}{(1+\lambda z_6)^i}\cdot\frac{1}{(1+\lambda z_1)^{k-i}} \\
    z_{3}=\frac{1}{(1+\lambda z_4)^{i-1}}\cdot\frac{1}{(1+\lambda z_2)^{k-i+1}}\\
    z_{4}=\frac{1}{(1+\lambda z_3)^{i-1}}\cdot\frac{1}{(1+\lambda z_7)^{k-i+1}}\\
    z_{5}=\frac{1}{(1+\lambda z_6)^{i-1}}\cdot\frac{1}{(1+\lambda z_1)^{k-i+1}}\\
    z_{6}=\frac{1}{(1+\lambda z_5)^{i-1}}\cdot\frac{1}{(1+\lambda z_8)^{k-i+1}}\\
    z_{7}=\frac{1}{(1+\lambda z_5)^i}\cdot\frac{1}{(1+\lambda z_8)^{k-i}} \\
    z_{8}=\frac{1}{(1+\lambda z_3)^i}\cdot\frac{1}{(1+\lambda z_7)^{k-i}}. \\
\end{array}%
\right.\eqno(2)
$$
╧хЁхяш°хь ёшёЄхьє єЁртэхэшщ (2) т тшфх
$$
\left\{%
\begin{array}{ll}
    z_{1}=(\frac{1+\lambda z_2}{1+\lambda z_4})^i\cdot\frac{1}{(1+\lambda z_2)^{k}} \\
    z_{2}=(\frac{1+\lambda z_1}{1+\lambda z_6})^i\cdot\frac{1}{(1+\lambda z_1)^{k}} \\
    z_{3}=(\frac{1+\lambda z_2}{1+\lambda z_4})^{i-1}\cdot\frac{1}{(1+\lambda z_2)^{k}}\\
    z_{4}=(\frac{1+\lambda z_7}{1+\lambda z_3})^{i-1}\cdot\frac{1}{(1+\lambda z_7)^{k}}\\
    z_{5}=(\frac{1+\lambda z_1}{1+\lambda z_6})^{i-1}\cdot\frac{1}{(1+\lambda z_1)^{k}} \\
    z_{6}=(\frac{1+\lambda z_8}{1+\lambda z_5})^{i-1}\cdot\frac{1}{(1+\lambda z_8)^{k}} \\
    z_{7}=(\frac{1+\lambda z_8}{1+\lambda z_5})^i\cdot\frac{1}{(1+\lambda z_8)^{k}}\\
    z_{8}=(\frac{1+\lambda z_7}{1+\lambda z_3})^i\cdot\frac{1}{(1+\lambda z_7)^{k}}.\\
\end{array}%
\right.\eqno(3)
$$
╨рчфхышт т ¤Єющ ёшёЄхьх єЁртэхэшщ яхЁтюх єЁртэхэшх эр ЄЁхЄ№х,
тЄюЁюх эр я Єюх, °хёЄюх эр ёхф№ьюх, ўхЄтхЁЄюх эр тюё№ьюх, яюыєўшь
ёыхфє■∙є■ ёшёЄхьє єЁртэхэшщ:
\\
$$
\left\{%
\begin{array}{ll}
    \frac{z_{1}}{z_{3}}=\frac{1+\lambda z_2}{1+\lambda z_4}\\
    \frac{z_{2}}{z_{5}}=\frac{1+\lambda z_1}{1+\lambda z_6}\\
    \frac{z_{6}}{z_{7}}=\frac{1+\lambda z_5}{1+\lambda z_8}\\
    \frac{z_{4}}{z_{8}}=\frac{1+\lambda z_3}{1+\lambda z_7}.\\
\end{array}%
\right. $$ ╚ёяюы№чє  ¤Єє ёшёЄхьє єЁртэхэшщ, (3) ьюцэю яхЁхяшёрЄ№
ъръ:

$$
\left\{%
\begin{array}{ll}
    z_{1}=(\frac{z_1}{z_3})^i\cdot\frac{1}{(1+\lambda z_2)^{k}} \\
    z_{2}=(\frac{z_2}{z_5})^i\cdot\frac{1}{(1+\lambda z_1)^{k}} \\
    z_{3}=(\frac{z_1}{z_3})^{i-1}\cdot\frac{1}{(1+\lambda z_2)^{k}}\\
    z_{4}=(\frac{z_8}{z_4})^{i-1}\cdot\frac{1}{(1+\lambda z_7)^{k}}\\
    z_{5}=(\frac{z_2}{z_5})^{i-1}\cdot\frac{1}{(1+\lambda z_1)^{k}} \\
    z_{6}=(\frac{z_7}{z_6})^{i-1}\cdot\frac{1}{(1+\lambda z_8)^{k}} \\
    z_{7}=(\frac{z_7}{z_6})^i\cdot\frac{1}{(1+\lambda z_8)^{k}}\\
    z_{8}=(\frac{z_8}{z_4})^i\cdot\frac{1}{(1+\lambda z_7)^{k}}.\\
\end{array}%
\right.\eqno(4)
$$
╚ч яхЁтюую єЁртэхэш  ёшёЄхь√ (4) эрщфхь $z_3$, шч тЄюЁюую $z_5$,
шч ёхф№ьюую $z_6$, шч тюё№ьюую $z_4$ ш яюфёЄртшт шї т тюё№ьюх,
ёхф№ьюх, тЄюЁюх ш яхЁтюх єЁртэхэш  ёшёЄхь√ єЁртэхэшщ (2),
ёююЄтхЄёЄтхээю, яюыєўшь:
$$
\left\{%
\begin{array}{ll}
    z_{1}=\frac{(1+\lambda z_{7})^k}{((1+\lambda z_7)^{k/i}+\lambda z_8^{1-1/i})^i}\cdot\frac{1}{(1+\lambda z_2)^{k-i}} \\
    \\
    z_{2}=\frac{(1+\lambda z_8)^k}{((1+\lambda z_8)^{k/i}+\lambda z_7^{1-1/i})^i}\cdot\frac{1}{(1+\lambda z_1)^{k-i}} \\
    \\
    z_{7}=\frac{(1+\lambda z_1)^k}{((1+\lambda z_1)^{k/i}+\lambda z_2^{1-1/i})^i}\cdot\frac{1}{(1+\lambda z_8)^{k-i}} \\
    \\
    z_{8}=\frac{(1+\lambda z_2)^k}{((1+\lambda z_2)^{k/i}+\lambda z_1^{1-1/i})^i}\cdot\frac{1}{(1+\lambda z_7)^{k-i}} \\
\end{array}%
\right.\eqno(5)
$$
╨рёёьюЄЁшь юЄюсЁрцхэшх $W:R^4 \rightarrow R^4,$ юяЁхфхыхээюх
ёыхфє■∙шь юсЁрчюь:

$$
\left\{%
\begin{array}{ll}
    z_{1}^{'}=\frac{(1+\lambda z_{7})^k}{((1+\lambda z_7)^{k/i}+\lambda z_8^{1-1/i})^i}\cdot\frac{1}{(1+\lambda z_2)^{k-i}}
    \\[3mm]
    z_{2}^{'}=\frac{(1+\lambda z_8)^k}{((1+\lambda z_8)^{k/i}+\lambda z_7^{1-1/i})^i}\cdot\frac{1}{(1+\lambda z_1)^{k-i}}
    \\[3mm]
    z_{7}^{'}=\frac{(1+\lambda z_1)^k}{((1+\lambda z_1)^{k/i}+\lambda z_2^{1-1/i})^i}\cdot\frac{1}{(1+\lambda z_8)^{k-i}}
    \\[3mm]
    z_{8}^{'}=\frac{(1+\lambda z_2)^k}{((1+\lambda z_2)^{k/i}+\lambda z_1^{1-1/i})^i}\cdot\frac{1}{(1+\lambda z_7)^{k-i}} \\
\end{array}%
\right.\eqno(6)
$$

╟рьхЄшь, ўЄю (5) хёЄ№ єЁртэхэшх $z=W(z).$ ╫Єюс√ Ёх°шЄ№ ёшёЄхьє
єЁртэхэшщ (5), эрфю эрщЄш эхяюфтшцэ√х Єюўъш юЄюсЁрцхэш 
$z^{'}=W(z).$\

\textbf{╦хььр 1.}\textit{ ╬ЄюсЁрцхэшх $W$ шьххЄ шэтрЁшрэЄэ√х
ьэюцхёЄтр ёыхфє■∙шї тшфют:}
$$I_1=\{(z_1, z_2, z_7, z_8) \in R^4: z_1=z_2=z_7=z_8\}, \ \ I_2=\{(z_1, z_2, z_7, z_8)\in R^4: z_1=z_7, \ z_2=z_8\},$$
$$I_3=\{(z_1, z_2, z_7, z_8) \in R^4: z_1=z_2, z_7=z_8\}, \ \ I_4=\{(z_1, z_2, z_7, z_8)\in R^4: z_1=z_8, \ z_2=z_7\}.$$

\textbf{─юърчрЄхы№ёЄтю.} ╧юърцхь шэтрЁшрэЄэюёЄ№ $I_2$
(шэтрЁшрэЄэюёЄ№ $I_i, i=1,3,4$ фюърч√трхЄё  рэрыюушўэю). ▀ёэю, ўЄю
фы  ы■сюую $z^{*}=(z_1^{*}, z_2^{*}, z_7^{*}, z_8^{*})\in I_2$
шьххЄ ьхёЄю $z_1^{*}=z_7^{*}, \ z_2^{*}=z_8^{*}.$ ╬Єё■фр ш шч (6)
шьххь
$$
\left\{%
\begin{array}{ll}
    z_{1}^{'}=\frac{(1+\lambda z_{1}^{*})^k}{[(1+\lambda z_1^{*})^{k/i}+\lambda (z_2^{*})^{1-1/i}]^i}\cdot\frac{1}{(1+\lambda z_2^{*})^{k-i}}
    \\[3mm]
    z_{2}^{'}=\frac{(1+\lambda z_2^{*})^k}{[(1+\lambda z_2^{*})^{k/i}+\lambda (z_1^{*})^{1-1/i}]^i}\cdot\frac{1}{(1+\lambda z_1^{*})^{k-i}}
    \\[3mm]
    z_{7}^{'}=\frac{(1+\lambda z_1^{*})^k}{[(1+\lambda z_1^{*})^{k/i}+\lambda (z_2^{*})^{1-1/i}]^i}\cdot\frac{1}{(1+\lambda z_2^{*})^{k-i}}
    \\[3mm]
    z_{8}^{'}=\frac{(1+\lambda z_2^{*})^k}{[(1+\lambda z_2^{*})^{k/i}+\lambda (z_1^{*})^{1-1/i}]^i}\cdot\frac{1}{(1+\lambda z_1^{*})^{k-i}} \\
\end{array}%
\right.$$ Є.х. $z_1^{'}=z_7^{'}; z_2^{'}=z_8^{'},$ р ¤Єю чэрўшЄ
$z^{'}=W(z^{*})\in I_2.$\

\textbf{╟рьхўрэшх 1.} ╚ч юяЁхфхыхэшщ 2 ш 3 ёыхфєхЄ, ўЄю т ёыєўрх
$I_2$ (шыш $I_3$ шыш $I_4$) ёырсю яхЁшюфшўхёър  ьхЁр ├шссёр эх
ёютярфрхЄ яхЁшюфшўхёъющ, хёыш шч єёыютшщ $z_1=z_7, \ z_2=z_8$ (шыш
$z_1=z_2, \ z_7=z_8$ шыш $z_1=z_8, \ z_2=z_7$) т√ЄхърхЄ, ўЄю їюЄ 
с√ юфэю шч ЁртхэёЄт $z_1=z_3, z_2=z_5, z_4=z_8, z_6=z_7$ эх
т√яюыэ хЄё , Є.х. чэрўхэшх $z_i$ чртшёшЄ юЄ $x_{\downarrow}$.

\textbf{╦хььр 2.} \textit{┼ёыш эр шэтрЁшрэЄэ√ї ьэюцхёЄтрї $I_2,
I_3, I_4$ ёє∙хёЄтє■Є ёырсю яхЁшюфшўхёъшх ьхЁ√ ├шссёр, Єю юэш
 ты ■Єё  ышсю ЄЁрэёы Ўшюээю-шэтрЁшрэЄэ√ьш, ышсю ёырсю
яхЁшюфшўхёъшьш (эх яхЁшюфшўхёъшьш).}

\textbf{─юърчрЄхы№ёЄтю.} ╧ЁютхЁшь фы  $I_2$(юёЄры№э√х фюърч√тр■Єё 
рэрыюушўэю.) ╧єёЄ№ $z_1=z_7, \ z_2=z_8$. ╥юуфр шч ёшёЄхь√
єЁртэхэшщ (2) яЁш $z_2\neq z_4$ яюыєўшь
$$z_{1}=\frac{1}{(1+\lambda z_4)^i}\cdot\frac{1}{(1+\lambda z_2)^{k-i}}\neq z_{3}=\frac{1}{(1+\lambda z_4)^{i-1}}\cdot\frac{1}{(1+\lambda
z_2)^{k-i+1}},$$ р Єю, ўЄю $z_2\neq z_4$ ьюцэю єтшфхЄ№ шч тЄюЁюую
ш ўхЄтхЁЄюую єЁртэхэшщ Єющ цх ёшёЄхь√ (2).\

\textbf{╤ыєўрщ $I_2$.} ╧хЁхяш°хь ёшёЄхьє (5) эр $I_2$ яЁш $k=2,
i=1$
$$
\left\{%
\begin{array}{ll}
    z_{1}=\frac{(1+\lambda z_{1})^2}{(1+\lambda z_1)^{2}+\lambda} \cdot\frac{1}{1+\lambda z_2}
    \\[3mm]
    z_{2}=\frac{(1+\lambda z_2)^2}{(1+\lambda z_2)^{2}+\lambda }\cdot\frac{1}{1+\lambda z_1} \\
    \end{array}%
\right.\eqno(7)
$$
╧юёых юсючэрўхэшщ $x=1+\lambda z_1$ ш $y=1+\lambda z_2$ шч ёшёЄхь√
(7) шьххь

$$
\left\{%
\begin{array}{ll}
    x=f(y)    \\
    y=f(x),
    \end{array}%
\right.
$$ уфх $f(x)={\lambda x^2\over(x^2+\lambda)(x-1)}$. ╨рёёьюЄЁшь яЁюшчтюфэє■
$$f'(x)=-{x\lambda (x^3-\lambda x+2\lambda)\over (x^2+\lambda)^2(x-1)^2}$$
ш яю ЇюЁьєых ╩рЁфрэю эрщфхь ъюЁэш ьэюуюўыхэр $x^3-\lambda
x+2\lambda$. ▌ЄюЄ ьэюуюўыхэ шьххЄ юфшэ тх∙хёЄтхээ√щ юЄЁшЎрЄхы№э√щ
ъюЁхэ№
$$x={1\over 3}\sqrt[3]{-27\lambda+3\lambda\sqrt{81-3\lambda}}+{\lambda\over \sqrt[3]{-27\lambda+3\lambda\sqrt{81-3\lambda}}}<0$$
яЁш $\lambda\leq 27$. ╟эрўшЄ, яЁш ¤Єюь єёыютшш $f'(x)<0$ яЁш
$x>1$, Є.х. ЇєэъЎш  $f(x)$ єс√трхЄ эр ¤Єюь шэЄхЁтрых ш єЁртэхэшх
$f(x)=x$ шьххЄ хфшэёЄтхээюх Ёх°хэшх. ┬ююс∙х уютюЁ , єЁртэхэшх
$f(x)=x$ шьххЄ хфшэёЄтхээюх Ёх°хэшх яЁш ы■с√ї $\lambda>0$, Є.ъ.
єЁртэхэшх
$$x={\lambda x^2\over(x^2+\lambda)(x-1)}=f(x)$$
¤ътштрыхэЄэю єЁртэхэш■ $x^3-x^2-\lambda=0$, ъюЄюЁюх яю шчтхёЄэющ
ЄхюЁхьх ю ъюышўхёЄтх яюыюцшЄхы№э√ї ъюЁэхщ ьэюуюўыхэр шьххЄ эх
сюыхх юфэюую яюыюцшЄхы№эюую Ёх°хэш , яюЄюьє ўЄю чэръш яЁш
ъю¤ЇЇшЎшхэЄрї ьхэ ■Єё  Єюы№ъю юфшэ Ёрч (╨шё.1).

\begin{center}
\includegraphics[width=10cm]{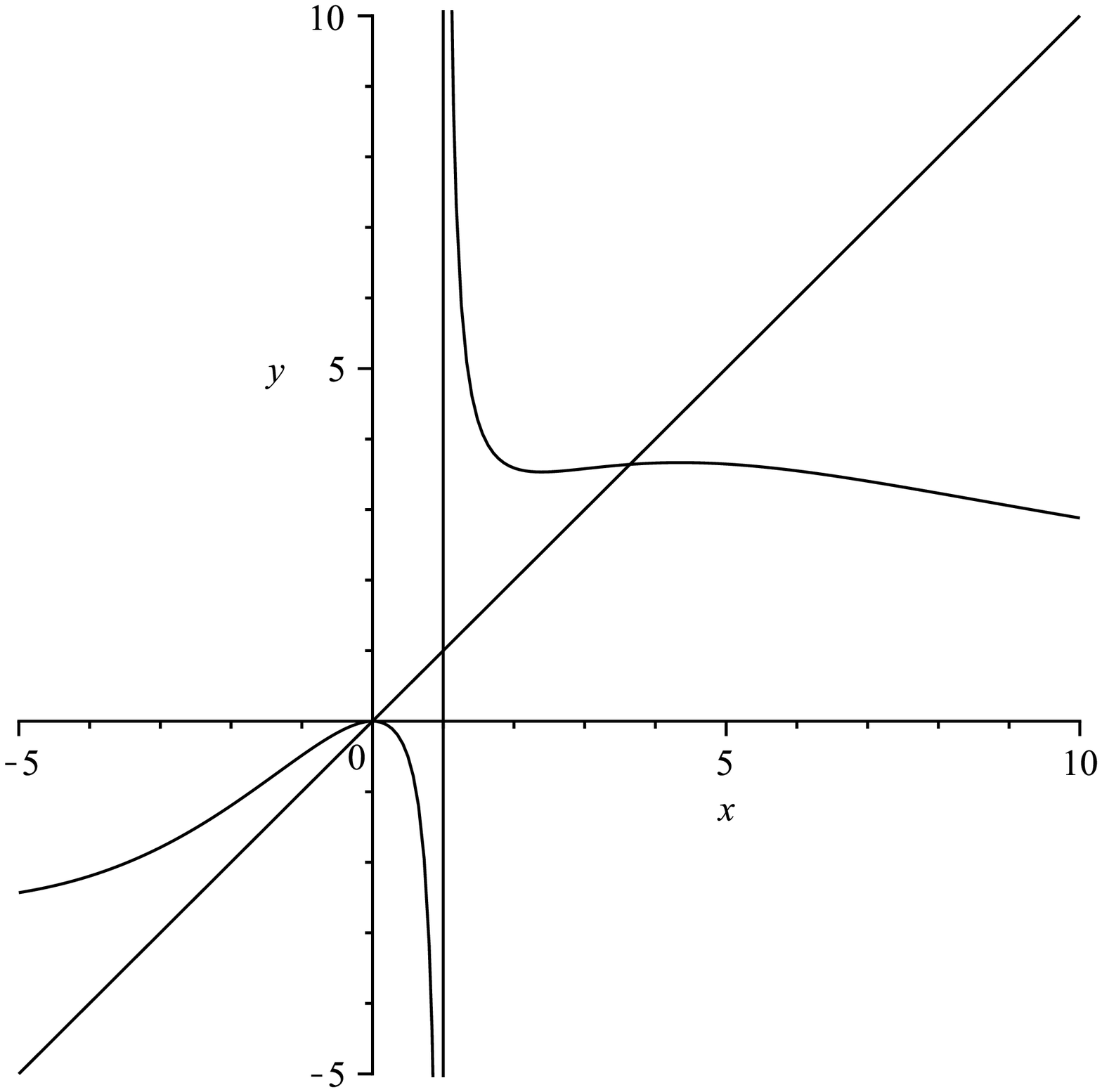}
\end{center}
\begin{center}{\footnotesize \noindent
 ╨шё.~1.
 ├ЁрЇшъш ЇєэъЎшщ $y=x$ ш $y=f(x)$ яЁш $\lambda=35$}
\end{center}

╬ўхтшфэю, ўЄю ¤Єю Ёх°хэшх сюы№°х хфшэшЎ√. ╩Ёюьх Єюую, юэю
эрїюфшЄё  ёЁхфш Ёх°хэшщ єЁртэхэш  $f(f(x))=x$. ╧ю¤Єюьє ЁрёёьюЄЁшь
єЁртэхэшх
$${x-f(f(x))\over x-f(x)}=0,$$
ъюЄюЁюх ¤ътштрыхэЄэю єЁртэхэш■
$$h(x)=x^6-(\lambda+2)x^5+(5\lambda+1)x^4-\lambda(2\lambda+5)x^3+2\lambda(2\lambda+1)x^2-3\lambda^2x+\lambda^2=0.$$
╚ч ¤Єюую єЁртэхэш  шьххь $h(1)=\lambda>0$ ш $h(x)\rightarrow
+\infty$ яЁш $x\rightarrow+\infty$. ╩Ёюьх Єюую, уЁрЇшъ ЇєэъЎшш
$h(x)$ ърёрхЄё  юёш Ox яЁш $x=2, \lambda=4$, Є.ъ.
$h(2)=-(\lambda-4)(5\lambda+4).$ ╬Єё■фр єЁртэхэшх $h(x)=0$ эх
шьххЄ Ёх°хэшщ яЁш $\lambda<4$, шьххЄ юфэю Ёх°хэшх яЁш $\lambda=4$
ш шьххЄ яю ъЁрщэхщ ьхЁх фтр Ёх°хэш  яЁш $\lambda>4$ (╨шё.2).

\begin{center}
\includegraphics[width=10cm]{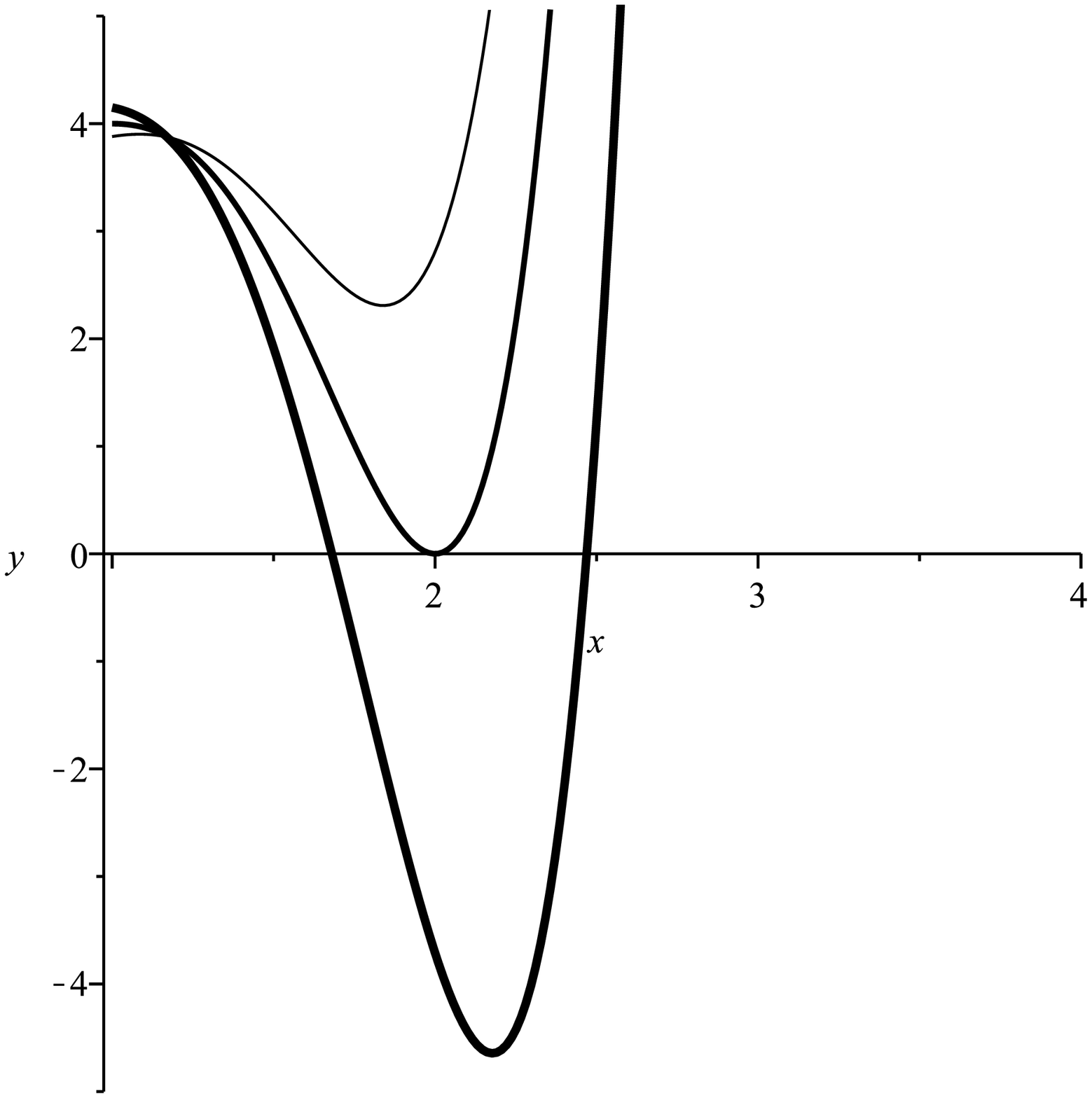}
\end{center}
\begin{center}{\footnotesize \noindent
 ╨шё.~2.
 ├ЁрЇшъ ЇєэъЎшш $h(x)$ яЁш $k=2$, $\lambda=3.88$ (ётхЁїє), яЁш ъЁшЄшўхёъющ $\lambda=4$ (яю ёхЁхфшэх) ш яЁш $\lambda=4.15$ (ёэшчє)}
\end{center}

╚Єръ, тхЁэю ёыхфє■∙хх

\textbf{╙ЄтхЁцфхэшх 1.} \textit{╤шёЄхьр єЁртэхэшщ (7) шьххЄ Єюы№ъю
юфэю Ёх°хэшх яЁш $\lambda<4$, шьххЄ фтр Ёх°хэш  яЁш $\lambda=4$ ш
шьххЄ эх ьхэхх ЄЁхї Ёх°хэшщ яЁш
$\lambda>4$.}\\

\textbf{╟рьхўрэшх 2.} ╤ яюью∙№■ ъюья№■ЄхЁэюую рэрышчр ьюцэю
єтшфхЄ№, ўЄю (7) шьххЄ Єюы№ъю ЄЁш Ёх°хэш  яЁш $\lambda>4$ (╨шё.2).

╬сючэрўр  $x=1+\lambda z_1$, $y=1+\lambda z_2$, ЁрёёьюЄЁшь ёшёЄхьє
(5) эр $I_2$ яЁш $k=3, i=1$
$$
\left\{%
\begin{array}{ll}
    y^2=\frac{\lambda x^3}{(x^3+\lambda)(x-1)}   \\[3mm]
    x^2=\frac{\lambda y^3}{(y^3+\lambda)(y-1)}   \\
    \end{array}%
\right.\eqno(8)
$$ ┬ ¤Єющ ёшёЄхьх єЁртэхэшщ эрщфхь шч яхЁтюую єЁртэхэш  $y$ ш яюфёЄртшь тю тЄюЁюх
єЁртэхэшх. ╥юуфр яюыєўшь єЁртэхэшх
$$f(x,\lambda)=x^{16}-(\lambda+4)x^{15}+3(\lambda+2)x^{14}-4x^{13}+(1-14\lambda)x^{12}+3\lambda(\lambda+8)x^{11}-$$
$$16\lambda x^{10}-4\lambda(5\lambda-1)x^9+36\lambda^2x^8+\lambda^2(\lambda-24)x^7+\lambda^2(6-13\lambda)x^6+$$
$$24\lambda^3x^5-16\lambda^3x^4+\lambda^3(4-3\lambda)x^3+6\lambda^4x^2-4\lambda^4x+\lambda^4=0,$$
фы  ъюЄюЁюую $f(1,\lambda)=\lambda^2>0, \ f(1.5,1.8)=-0.5255524<0,
\ f(1.8,1.8)=9.30017>0, \ f(2,1.8)=-90.1232<0$ ш
$f(x,\lambda)\rightarrow +\infty$ яЁш $x\rightarrow+\infty, \
\lambda>0.$ ╤ыхфютрЄхы№эю, яюёыхфэхх єЁртэхэшх шьххЄ эх ьхэхх
ўхЄ√Ёхї Ёх°хэшщ яЁш $x>1$ (╨шё.3).

\begin{center}
\includegraphics[width=10cm]{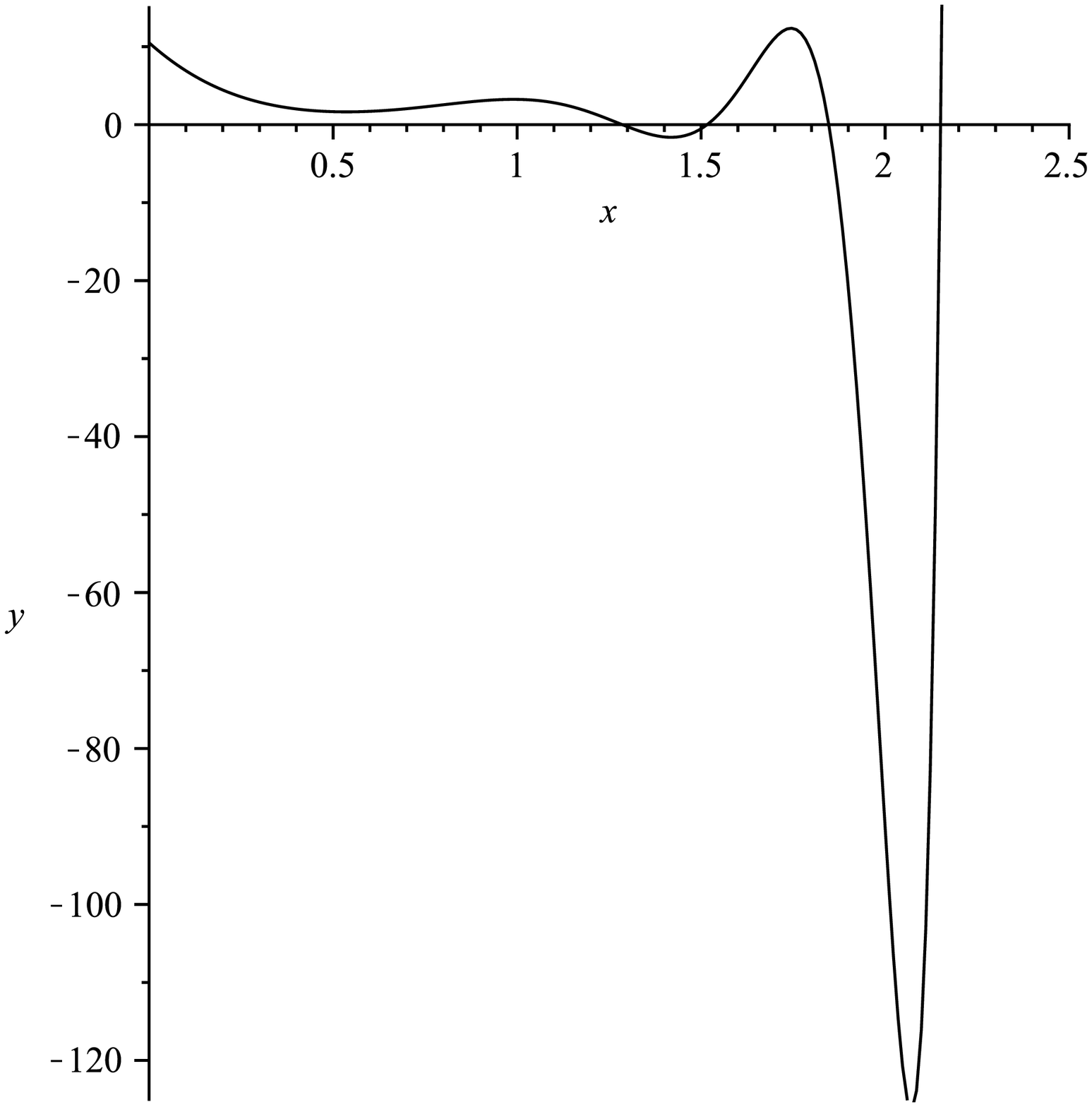}
\end{center}
\begin{center}{\footnotesize \noindent
 ╨шё.~3.
 ├ЁрЇшъ ЇєэъЎшш $f(x,\lambda)$ яЁш $\lambda=1.8$}
\end{center}

\textbf{╙ЄтхЁцфхэшх 2.} \textit{╤є∙хёЄтєхЄ Єрър  $\lambda_{0}$,
ўЄю ёшёЄхьр єЁртэхэшщ (8) шьххЄ эх ьхэхх ўхЄ√Ёхї Ёх°хэшщ яЁш
$\lambda>\lambda_{0}$.}\

\textbf{╟рьхўрэшх 3.} ╤ яюью∙№■ ъюья№■ЄхЁэюую рэрышчр ьюцэю
єтшфхЄ№, ўЄю ёє∙хёЄтєхЄ $\lambda_{cr}$, Єрър , ўЄю (8) шьххЄ фтр
Ёх°хэш  яЁш $\lambda<\lambda_{cr}$ ш ўхЄ√Ёх Ёх°хэш  яЁш
$\lambda>\lambda_{cr}$. ╧Ёш ¤Єюь т ърцфюь ёыєўрх юфэю шч Ёх°хэшщ
ёююЄтхЄёЄтєхЄ ЄЁрэёы Ўшюээю-шэтрЁшрэЄэющ ьхЁх ├шссёр.

\textbf{╤ыєўрщ $I_3$.} ╧хЁхяш°хь ёшёЄхьє (5) эр $I_3$ яЁш $k\geq1,
i=1$
$$
\left\{%
\begin{array}{ll}
    z_{1}=\frac{(1+\lambda z_{7})^k}{(1+\lambda z_7)^{k}+\lambda} \cdot\frac{1}{(1+\lambda z_1)^{k-1}}
    \\[3mm]
    z_{7}=\frac{(1+\lambda z_1)^k}{(1+\lambda z_1)^{k}+\lambda }\cdot\frac{1}{(1+\lambda z_7)^{k-1}} \\
    \end{array}%
\right.\eqno(9)
$$
╧юёых юсючэрўхэшщ $x=1+\lambda z_1$ ш $y=1+\lambda z_7$ шч ёшёЄхь√
(9) шьххь

$$
\left\{%
\begin{array}{ll}
    x^k-x^{k-1}={\lambda y^k\over y^k+\lambda}    \\
    y^k-y^{k-1}={\lambda x^k\over x^k+\lambda}.
    \end{array}%
\right.$$ ┬√ўшЄр  шч яхЁтюую єЁртэхэш  тЄюЁюх т ¤Єющ ёшёЄхьх,
яюыєўшь єЁртэхэшх
$$(x-y)[(x^{k-1}+...+y^{k-1}-(x^{k-2}+...+y^{k-2}))(x^k+\lambda)(y^k+\lambda)+\lambda^2(x^{k-1}+...+y^{k-1})]=0,$$
ъюЄюЁюх шьххЄ Єюы№ъю Ёх°хэшх тшфр $x=y$ яЁш $x>1, y>1.$ ╟рьхЄшь,
ўЄю хёыш $x=y$, Єю $z_1=z_2=z_3=z_4=z_5=z_6=z_7=z_8$ яЁш $i=1$.
─хщёЄтшЄхы№эю, яєёЄ№ $z_1=z_2=z_7=z_8$. ╥юуфр шч яхЁтюую ш тЄюЁюую
єЁртэхэшщ (2) яюыєўшь $z_4=z_6$, р шч ёхф№ьюую ш тюё№ьюую
єЁртэхэшщ Єющ цх ёшёЄхь√ (2) $z_3=z_5$. ─рыхх, шч ЄЁхЄ№хую ш
ўхЄтхЁЄюую єЁртэхэшщ шьххь
$$z_{3}=\frac{1}{(1+\lambda z_4)^{i-1}}\cdot\frac{1}{(1+\lambda
z_1)^{k-i+1}}, \   z_{4}=\frac{1}{(1+\lambda
z_3)^{i-1}}\cdot\frac{1}{(1+\lambda z_1)^{k-i+1}},$$ юЄъєфр
ёыхфєхЄ, ўЄю $z_3=z_4$ яЁш $i=1$. ╟эрўшЄ, $z_3=z_4=z_5=z_6.$
╤ыхфютрЄхы№эю, (2) ьюцэю яхЁхяшёрЄ№
$$
\left\{%
\begin{array}{ll}
    z_1=\frac{1}{(1+\lambda z_3)^i}\cdot\frac{1}{(1+\lambda z_1)^{k-i}} \\
    z_3=\frac{1}{(1+\lambda z_3)^{i-1}}\cdot\frac{1}{(1+\lambda z_1)^{k-i+1}}. \\
\end{array}%
\right.
$$ ╨рчфхышт яхЁтюх єЁртэхэшх эр тЄюЁюх т ¤Єющ ёшёЄхьх яюыєўшь, ўЄю
$z_1=z_3.$

\textbf{╙ЄтхЁцфхэшх 3.} \textit{╧єёЄ№ $k\geq1, i=1$. ╥юуфр ёшёЄхьр
єЁртэхэшщ (9) шьххЄ хфшэёЄтхээюх Ёх°хэшх яЁш $\lambda>0$.}\\

\textbf{╤ыєўрщ $I_4$.} ╧хЁхяш°хь ёшёЄхьє (5) эр $I_4$ яЁш $k\geq1,
i=1$
$$
\left\{%
\begin{array}{ll}
    z_{1}=\frac{1+\lambda z_{2}}{(1+\lambda z_2)^{k}+\lambda} \\[3mm]
    z_{2}=\frac{1+\lambda z_1}{(1+\lambda z_1)^{k}+\lambda } \\
    \end{array}%
\right.\eqno(10)
$$
╧юёых юсючэрўхэшщ $x=1+\lambda z_1$ ш $y=1+\lambda z_2$ шч ёшёЄхь√
(10) шьххь

$$
\left\{%
\begin{array}{ll}
    x=f(y)    \\
    y=f(x),
    \end{array}%
\right.
$$ уфх $f(x)={\lambda x\over x^k+\lambda}+1$. ╟рьхЄшь, ўЄю $f(0)=1, f(1)={\lambda\over
1+\lambda}+1>1.$ ┬√ўшёышь яЁюшчтюфэє■
$$f'(x)=-\lambda{(k-1)x^k-\lambda\over (x^k+\lambda)^2}.$$
╬Єё■фр ЇєэъЎш  $f(x)$ тючЁрёЄрхЄ яЁш $0<x<\sqrt[k]{{\lambda\over
k-1}}$ ш єс√трхЄ яЁш $x>\sqrt[k]{{\lambda\over k-1}}$, Є.х.
$x_{max}=\sqrt[k]{{\lambda\over k-1}}.$ ╚ч тёхую ёърчрээюую
ёыхфєхЄ, ўЄю єЁртэхэшх $f(x)=x$ шьххЄ хфшэёЄтхээюх Ёх°хэшх яЁш
$x>1, \lambda>0$. ▌Єю х∙х ьюцэю єтшфхЄ№, рэрыюушўэю ёыєўр■ $I_2$,
шёяюы№чє  ЄхюЁхьє ю ъюышўхёЄтх яюыюцшЄхы№э√ї ъюЁэхщ ьэюуюўыхэр
(╨шё.4).

\begin{center}
\includegraphics[width=10cm]{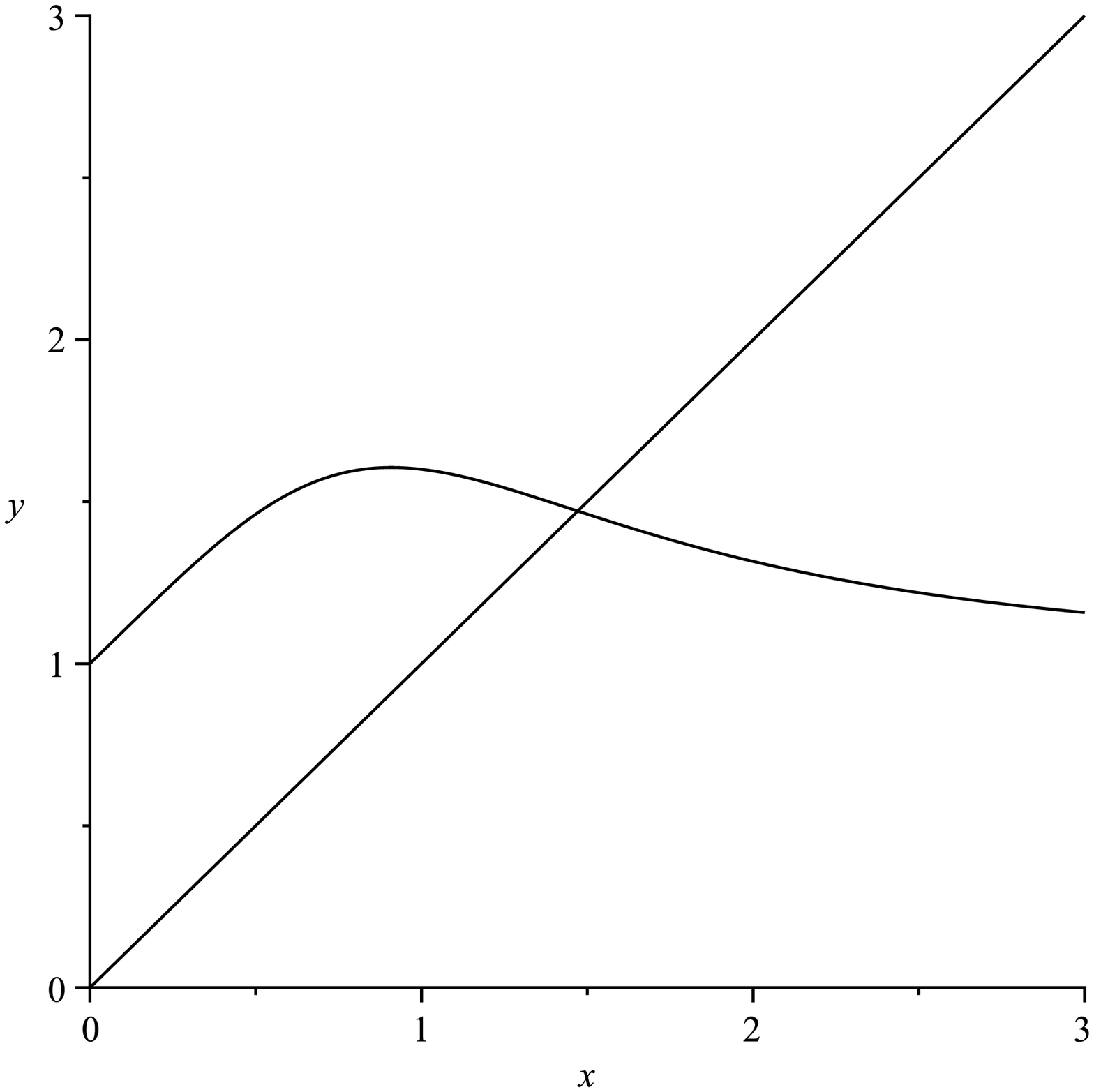}
\end{center}
\begin{center}{\footnotesize \noindent
 ╨шё.~4.
 ├ЁрЇшъш ЇєэъЎшщ $y=x$ ш $y=f(x)$ яЁш $k=3$, $\lambda=1.5$}
\end{center}

└эрыюушўэю ёыєўр■ $I_2$ яЁш $k=2, i=1$ ш $k=3, i=1$ яюыєўшь
$$h_1(x)=(\lambda+1)x^2+\lambda x+2\lambda^2+\lambda,$$
$$h_2(x)=(\lambda+1)x^6-\lambda x^5+2\lambda x^4+2\lambda(\lambda+1)x^3+2\lambda^2 x+2\lambda^3+\lambda^2,$$
ёююЄтхЄёЄтхээю. ╬ўхтшфэю, ўЄю $h_1(x)>0,$ $h_2(x)>0$ яЁш $x>1,
\lambda>0$. ╟эрўшЄ, (10) эх шьххЄ Ёх°хэшщ, ъЁюьх $z_1=z_2$, Є.х.
ёяЁртхфыштю

\textbf{╙ЄтхЁцфхэшх 4.} ╤шёЄхьр єЁртэхэшщ (10) шьххЄ хфшэёЄтхээюх
Ёх°хэшх яЁш $k=2$ ш $k=3$.

┬ ёшыє тёхї єЄтхЁцфхэшщ ш ╦хьь√ 2 ёяЁртхфыштр ёыхфє■∙р 

\textbf{╥хюЁхьр.} \textit{─ы  HC-ьюфхыш т ёыєўрх эюЁьры№эюую
фхышЄхы  шэфхъёр ўхЄ√Ёх тхЁэ√ ёыхфє■∙шх єЄтхЁцфхэш : }

\textit{1. ╧Ёш $k\geq1, i\leq k$ эр $I_1$ ёырсю яхЁшюфшўхёър  ьхЁр
├шссёр хфшэёЄтхээр. ┴юыхх Єюую, ¤Єр ьхЁр ёютярфрхЄ ё хфшэёЄтхээющ
ЄЁрэёы Ўшюээю-шэтрЁшрэЄэющ ьхЁющ ├шссёр.}

\textit{2. ╧єёЄ№ $k=2, i=1, \lambda_{cr}=4$. ╥юуфр эр $I_2$ яЁш
$\lambda<\lambda_{cr}$ ёє∙хёЄтєхЄ юфэр ёырсю яхЁшюфшўхёър  ьхЁр
├шссёр, ъюЄюЁр   ты хЄё  ЄЁрэёы Ўшюээю-шэтрЁшрэЄэющ, яЁш
$\lambda=\lambda_{cr}$ ёє∙хёЄтє■Є фтх ёырсю яхЁшюфшўхёъшх ьхЁ√
├шсёёр, юфэр шч ъюЄюЁ√ї  ты хЄё  ЄЁрэёы Ўшюээю-шэтрЁшрэЄэющ,
фЁєур  ёырсю яхЁшюфшўхёъющ (эх яхЁшюфшўхёъющ) ш яЁш
$\lambda>\lambda_{cr}$ ёє∙хёЄтє■Є эх ьхэхх фтєї ёырсю
яхЁшюфшўхёъшї (эх яхЁшюфшўхёъшї) ьхЁ ├шсёёр.}

\textit{3. ╧єёЄ№ $k=3, i=1$. ╥юуфр ёє∙хёЄтєхЄ $\lambda_0$ Єрър ,
ўЄю эр $I_2$ яЁш $\lambda>\lambda_{0}$ ёє∙хёЄтє■Є эх ьхэхх ўхЄ√Ёхї
ьхЁ ├шссёр, юфэр шч ъюЄюЁ√ї  ты хЄё  ЄЁрэёы Ўшюээю-шэтрЁшрэЄэющ, р
юёЄры№э√х ёырсю яхЁшюфшўхёъшьш (эх яхЁшюфшўхёъшьш) ьхЁрьш ├шссёр.}

\textit{4. ╧Ёш $k\geq1, i=1$ эр $I_3$ ёырсю яхЁшюфшўхёър  ьхЁр
├шссёр хфшэёЄтхээр.}

\textit{5. ╧Ёш $k=2,3, i=1$ эр $I_4$ ёырсю яхЁшюфшўхёър  ьхЁр
├шссёр
хфшэёЄтхээр.}\\

\textbf{╟рьхўрэшх 4.} ╩юья№■ЄхЁэ√щ рэрышч яюърч√трхЄ, ўЄю ёшёЄхьр
(10) шьххЄ Єюы№ъю юфэю Ёх°хэшх яЁш $k=4,5,6$ ш $\lambda>0$. └ яЁш
$k\geq 7$ яЁш эхъюЄюЁ√ї чэрўхэш ї $\lambda$ ьюцэю єтшфхЄ№, ўЄю є
ёшёЄхь√ (10) Ёх°хэшх эх хфшэёЄтхээю, Є.х. ёє∙хёЄтє■Є ёырсю
яхЁшюфшўхёъшх (эх яхЁшюфшўхёъшх) ьхЁ√ ├шссёр фы  HC-ьюфхыш
(╨шё.5).

\begin{center}
\includegraphics[width=10cm]{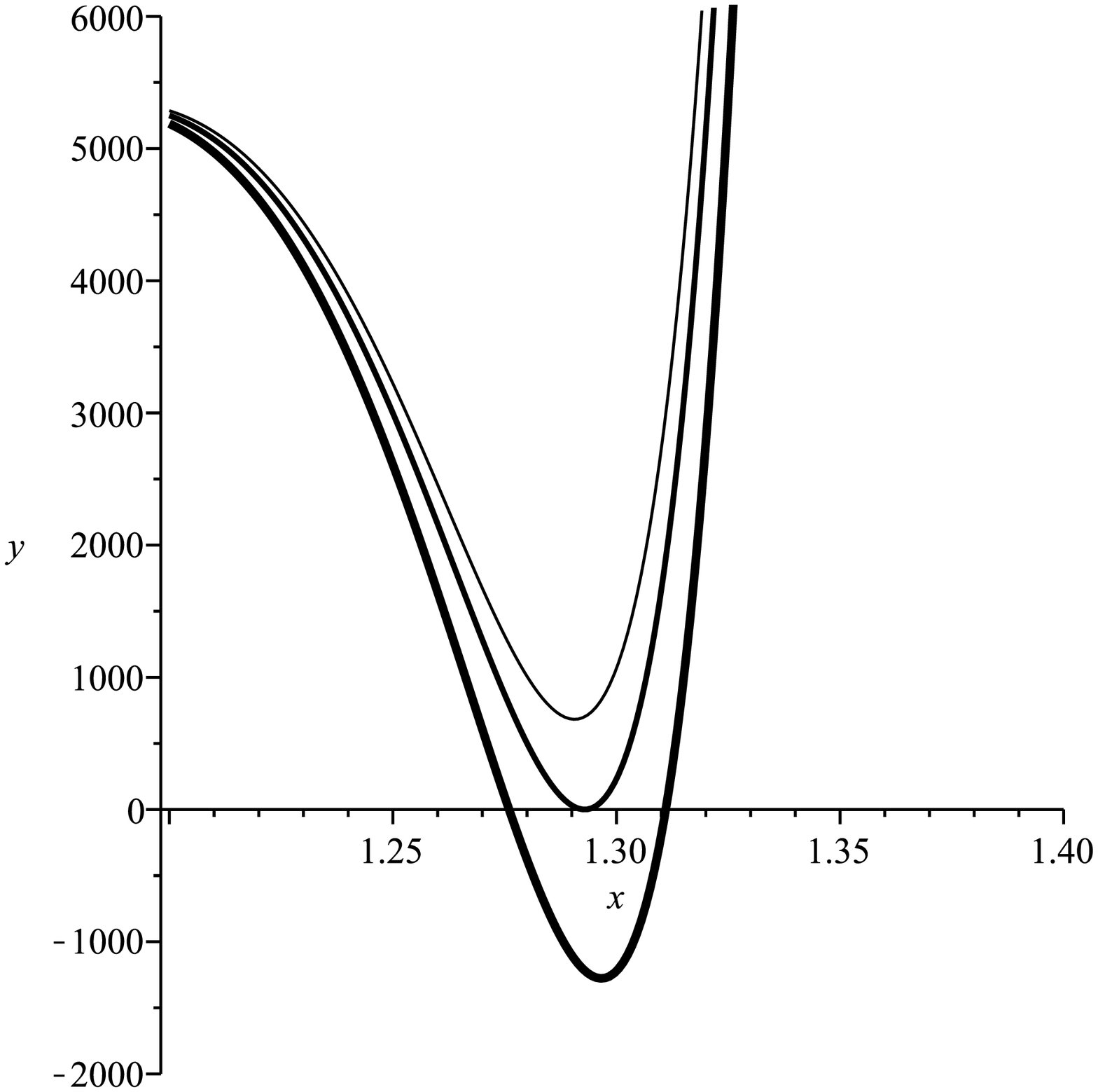}
\end{center}
\begin{center}{\footnotesize \noindent
 ╨шё.~5.
 ├ЁрЇшъ ЇєэъЎшш $h(x)$ яЁш $k=7$, $\lambda=1.765$ (ётхЁїє), яЁш $\lambda\approx 1.768674523476329362$ (яю ёхЁхфшэх) ш яЁш $\lambda=1.775$ (ёэшчє)}
\end{center}

\end{document}